\documentclass[12pt]{article}
\usepackage{epsf}
\usepackage{graphicx}
\usepackage{amssymb}
\def\hybrid{\topmargin 0pt      \oddsidemargin 0pt
	\headheight 0pt \headsep 0pt
	\textheight 9in         % US paper
	\textwidth 6.25in       % A4 paper
	\marginparwidth .875in
	\parskip 5pt plus 1pt   \jot = 1.5ex}

\catcode`\@=11
\def\marginnote#1{}
\newcount\hour
\newcount\minute
\newtoks\amorpm
\hour=\time\divide\hour by60
\minute=\time{\multiply\hour by60 \global\advance\minute by-\hour}
\edef\standardtime{{\ifnum\hour<12 \global\amorpm={am}%
	\else\global\amorpm={pm}\advance\hour by-12 \fi
	\ifnum\hour=0 \hour=12 \fi
	\number\hour:\ifnum\minute<10 0\fi\number\minute\the\amorpm}}
\edef\militarytime{\number\hour:\ifnum\minute<10 0\fi\number\minute}

\def\draftlabel#1{{\@bsphack\if@filesw {\let\thepage\relax
   \xdef\@gtempa{\write\@auxout{\string
      \newlabel{#1}{{\@currentlabel}{\thepage}}}}}\@gtempa
   \if@nobreak \ifvmode\nobreak\fi\fi\fi\@esphack}
	\gdef\@eqnlabel{#1}}
\def\@eqnlabel{}
\def\@vacuum{}
\def\draftmarginnote#1{\marginpar{\raggedright\scriptsize\tt#1}}

\def\draft{\oddsidemargin -.5truein
	\def\@oddfoot{\sl preliminary draft \hfil
	\rm\thepage\hfil\sl\today\quad\militarytime}
	\let\@evenfoot\@oddfoot \overfullrule 3pt
	\let\label=\draftlabel
\let\marginnote=\draftmarginnote
	\let\marginnote=\draftmarginnote
   \def\@eqnnum{(\theequation)\rlap{\kern\marginparsep\tt\@eqnlabel}%
\global\let\@eqnlabel\@vacuum}  }

\def\numberbysection{\@addtoreset{equation}{section}
	\def\theequation{\thesection.\arabic{equation}}}

\def\underline#1{\relax\ifmmode\@@underline#1\else
	$\@@underline{\hbox{#1}}$\relax\fi}

\def\titlepage{\@restonecolfalse\if@twocolumn\@restonecoltrue\onecolumn
     \else \newpage \fi \thispagestyle{empty}\c@page\z@
	\def\thefootnote{\fnsymbol{footnote}} }

\def\endtitlepage{\if@restonecol\twocolumn \else  \fi
	\def\thefootnote{\arabic{footnote}}
	\setcounter{footnote}{0}}  %\c@footnote\z@ }
\catcode`@=12
\relax

\def\beq{\begin{equation}}
\def\eeq{\end{equation}}
\def\bea{\begin{eqnarray}}
\def\eea{\end{eqnarray}}

\relax
\newcommand{\WD}[2]{W\!D_{#1}^{#2}}

\newcommand{\z}[2]{\mathbb{Z}^{(2)}_{#1}(#2)}
\newcommand{\ud}{\mathrm{d}}

\hybrid
\begin{document}

\begin{titlepage}
\begin{center}
December~2008 \hfill . \\[.5in]
{\large\bf Renormalization group flows for the second $\mathbb{Z}_{N}$ parafermionic field theory for $N$ even.} 
\\[.5in] 
{\bf Benoit Estienne}\\[.2in]
{\it LPTHE, CNRS, UPMC Univ Paris 06 }\\

 {\it Bo\^{\i}te 126, 4 place Jussieu, F-75252 Paris Cedex 05}\\
estienne@lpthe.jussieu.fr \\

\end{center}

\underline{Abstract.}

Extending the results obtained in the case $N$ odd, the effect of slightly relevant perturbations of the second parafermionic field theory with the symmetry $\mathbb{Z}_{N}$, for $N$ even, are studied. The renormalization group equations, and their infra red fixed points exhibit the same structure in both cases. In addition to the standard flow from the $p$-th to the $(p-2)$-th model, another fixed point corresponding to the $(p-1)$-th model is found.

\end{titlepage}
\newpage

Conformal field theories  describe critical points of some statistical system, i.e. the fixed point of the renomalization group. In order to get some insights into the neighborhood of this critical point, one can study the effects of slightly relevant perturbations of the correponding conformal field theory, using a renormalization group approach.

Parafermionic conformal field theories describe systems enjoying an extended conformal symmetry : they possess an additional cyclic symmetry $\mathbb{Z}_{N}$. While the first series of parafermionic conformal field theories \cite{ref1} are well studied and applied in various domains \cite{ref2,ref3,ref4}, far less is known about the second parafermionic series \cite{ref5,ref6,ref7,ref8}. These conformal theories have a richer structure than the first parafermions. For a given $N\geq 5$, there are infinitetely many second parafermionic theories $\z{N}{p}$, labeled by the integer $p \geq N-1$. These theories are unitary and correspond to degenerate representations of the corresponding parafermionic chiral algebra. The presence of the parameter $p$, for a given $\mathbb{Z}_{N}$, opens a way to reliable perturbative studies. It allows in particular to study the renormalisation group flows in the space of these conformal theory models, under various perturbations.

This problem has already been adressed for the case $N$ odd \cite{refletter,refpaper}, where the effect of two slightly relevant perturbations were studied. In this letter these results are generalized to the case $N$ even.

A particular case of this problem has already been treated. The second parafermionic theory $\z{N}{p}$ with $N$ even is a particular exemple of symmetric coset on a simply laced Lie algebra, and as such its perturbation by one particular relevant operator has been studied in \cite{ref16}. It is already known that the corresponding renormalization group equations describe a flow from $\z{N}{p}$ to $\z{N}{p-2}$.  Guided by the results obtained for the case $N$ odd, we consider a more general perturbation, and an additional fixed point is found.

The theory $\z{N}{p}$ perturbed by two appropriate fields will exhibit two different infrared (IR) fixed points :

\begin{itemize}

\item one of these fixed point is described by the expected $\z{N}{p-2}$

\item the other fixed point corresponds to $\z{N}{p-1}$

\end{itemize}

The details of the second $\mathbb{Z}_{N}$ parafermionic theory with $N$ even , $N \geq 6$ can be found in \cite{ref8}. The main difference between the odd and even cases lies in the coset construction of these CFT \cite{ref13}: 

\beq
\z{N}{p}=\frac{SO(N)_{k}\otimes SO(N)_{2}}{SO(N)_{k+2}} \quad p = k + N-2
\label{coset_ZN}
\eeq

Depending on the parity of $N$, the affine Lie algebra relevant to the coset construction will be either $D_r^{(1)}$, when $N=2r$ is even, or $B_r^{(1)}$, when $N=2r+1$. In both cases the central charge is :

\beq
c_N(p)= (N-1)\left(1-\frac{N(N-2)}{p(p+2)} \right)
\eeq

The chiral algebra is made of $N-1$ parafermionic currents $\Psi^k$, $k=1..N-1$ obeying the following fusion rules : 

\beq
\left[\Psi^k \right] \times \left[\Psi^l \right] \rightarrow \left[\Psi^{k+l} \right]
\label{algebra}
\eeq
  where the $\mathbb{Z}_N$ charges $k,l,k+l$ are defined modulo $N$.

The primary operators $\Phi_{(\vec{n}\mid\vec{n}')}$ are labeled by two vectors $\vec{n},\vec{n}'$ belonging to the weight lattice of the $X_r$ Lie algebra, where $X_r$ denotes either $B_r$ or $D_r$. Their conformal dimension is given by :

\beq
\Delta_{(\vec{n}\mid\vec{n}')} = \frac{\left( (p+2)\vec{n} - p \vec{n}'  \right)^2 - 4 {\vec{\rho}}^2}{2p(p+2)} + B_Q
\label{Kac}
\eeq

where $B_Q$ is the boundary term which depend on the sector of the theory, i.e. on the $\mathbb{Z}_N$ charge of the primary field, and  $\vec{\rho}= \sum_i \vec{\omega}_i$ is the Weyl vector of the corresponding Lie algebra.

When $N$ is even, the second parafermionic theory $\z{N}{p}$ is a coset of the form $G_k \otimes G_l / G_{k+l}$ on a simply laced Lie algebra $G = D_r$.  Fateev \cite{ref16} proved in that general framework that the perturbation by an appropriate operator $A$ : 

\beq
{\bf A}={\bf A}_{0} + g\int \ud^2 x \ A  (x)  
\eeq

is integrable, and leads for $g > 0$ to an IR fixed point described by $G_{k-l} \otimes G_l / G_{k}$, provided $k \gg 1$.

In terms of the GKO construction \cite{refGKO}, this appropriate operator field $A(x)$ correspond to the branching $[(k,id)\otimes(l,id)/(k+l,\ ad)]$, and has a conformal dimension :

\beq
\Delta_A = 1 - \frac{g}{k+l+g}
\eeq

$g$ being the Coxeter number of the simply laced Lie algebra $G$. For the second parafermionic theory $\z{N}{p}$ theories ($N$ even), this correspond to a $\mathbb{Z}_N$ neutral field with dimension $\Delta_A =  1-(N-2)/(p+2)$. If the ultraviolet CFT is described by $\z{N}{p}$, then the IR fixed point will correspond to $\z{N}{p-2}$.

In this letter a more general perturbation in considered, having in mind the results obtained for the odd case. Assuming $A_0$ to describe the second parafermionic field theory, the action of the perturbed CFT takes the following form : 

\beq
{\bf A}={\bf A}_{0} + \sum_{\alpha} g_{\alpha} \int \ud^2 x \ \Phi_{\alpha}  (x)  
\label{action}
\eeq

In order to preserve the $\mathbb{Z}_N$ symmetry, the relevant fields $\Phi_{\alpha}$ have to be neutral w.r.t. $\mathbb{Z}_N$. Perturbatively well controlled domain of $\z{N}{p}$ theories is that of $ p\gg 1$, giving a small parameter $\epsilon\sim 1/p$. This is similar to the original perturbative renormalization group treatment of minimal models for Virasoro algebra based conformal theory \cite{ref9,ref10}.

In this domain, i.e. for $p\gg 1$, one finds, in the lower part of the Kac table of $\z{N}{p}$ parafermionic theory, two $\mathbb{Z}_{N}$ neutral fields which are slightly relevant and which close by the operator algebra. They are:

\beq
S=\Phi_{(1 1 \ldots\mid 3 1 \ldots)}
\label{S}
\eeq

\beq
\begin{array}{ll}
 A  & = \left\{	\begin{array}{ll}
\Psi^{-1}_{-\frac{1}{3}}\Phi_{(1 1 1 \mid 1 2 2)} & \textrm{for }N=6\\
\Psi^{-1}_{-\frac{2}{N}}\Phi_{(1 1 1 \ldots \mid 1 2 1 \ldots)} & \textrm{for } N \geq 8
	\end{array} \right.
\end{array}
\label{A}
\eeq

The first one is a $\mathbb{Z}_{N}$ singlet and the second is a parafermionic algebra descendant of a doublet. They both belong to the neutral sector of $\mathbb{Z}_{N}$ and they are both Virasoro primaries.

Their labeling as $S$ and $A$ is just our shortened notations for these fields, and the field $A$ is precisely the operator Fateev used as a perturbation.

Their  conformal dimensions are given by (\ref{Kac}) :
\beq  \begin{array}{lll}
\Delta_{A}& = &1-(2r-2)\epsilon \\
\Delta_{S}& = & 1-2r\epsilon 
\end{array}
\eeq

where $\epsilon$ is defined as follows:

\beq
\epsilon=\frac{1}{p+2}\simeq\frac{1}{p}
\eeq

The fields $S$ and $A$ have the same structure as in the case $N$ odd \cite{refletter,refpaper}. Going back to the parameter $N$ rather than $r$, their conformal dimension for both cases , $N$ odd and even, read :

\beq
\begin{array}{lll}
\Delta_{A}&= &1-(N-2)\epsilon \\
\Delta_{S}&= &1-N\epsilon 
\end{array}
\label{relev_dim}
\eeq

Perturbing with the fields $S$ and $A$ corresponds to taking the action of the theory in the form:
\beq
{\bf A}={\bf A}_{0}+\frac{2g}{\pi}\int d^{2}xS(x)+\frac{2h}{\pi}\int d^{2}x A(x) 
\eeq
where $g$ and $h$ are the corresponding coupling constants; 
the additional factors  $\frac{2}{\pi}$  are added to simplify the coefficients 
of the renormalization group equations which follow;  we recall that ${\bf A}_{0}$ is assumed 
to be the action of the unperturbed $\z{N}{p}$ conformal theory.

The operator algebra of the fields $S$ and $A$ is of the form:
\beq 
S(x')S(x)=\frac{D_{1}}{|x'-x|^{4\Delta_{S}-2\Delta_{A}}}A(x)+...
\label{fSS}
\eeq
\beq
A(x')A(x)=\frac{D_{2}}{|x'-x|^{2\Delta_{A}}} A(x)+...
\label{fAA}
\eeq
\beq
S(x')A(x)=\frac{D_{1}}{|x'-x|^{2\Delta_{1}}} S(x)+...
\label{fSA}
\eeq

Only the fields which are relevant for the renormalization group flows are shown explicitly. The operator algebra expansions in (\ref{fSS})-(\ref{fSA}) and the constants $D_{1}$ and $D_{2}$ are obtained by the same fusion procedure as for the case $N$ odd \cite{refletter,refpaper}. From these operator product expansions one derives in a standard way the renormalization group equations for the couplings $g$ and $h$:
\bea
\beta_g = \frac{dg}{d\xi} & = & 2 (1-\Delta_S)  g - 4D_{1}g h  \label{betag} \\
\beta_h = \frac{dh}{d\xi} & = & 2 (1-\Delta_A)  h - 2D_{1}g^{2}-2D_{2}h^{2} \label{betah} 
\eea
These are up to (including) the first non-trivial order of the perturbations in $g$ and $h$. All this analysis is valid for $\epsilon \ll 1$, and the above RG flow equations are valid in the region $g,h \precsim \epsilon$.

These equations derive from a potential:

\begin{eqnarray}
\beta_g & = & \partial_{g} V\\ 
\beta_h & = & \partial_{h} V
\end{eqnarray}

with :

\begin{equation}
V(g,h)  =  (1-\Delta_S) g^2 + (1-\Delta_A) h^2 - 2/3 D_{2}h^{3} - 2D_{1}h g^{2} \label{potential}
\end{equation}

This potential plays a central role in the renormalization group flows. Let us consider the function $c(g,h)$ defined by $ c(g,h) = c_0 - \frac{V(g,h)}{24}$ : this is the c-function introduced by Zamolodchikov, which decreases along the renormalization group flows, and coincide with the central charge at any fixed point. In order to analyse the presence of IR fixed points for the renormalization group, and predict the corresponding central charges, explicit values for the operator algebra constants $D_1$ and $D_2$ are required, at least in the leading order in $\epsilon$ : they are obtained by a fusion procedure. This amounts to relate the second parafermionic theory $\z{N}{p}$ to $WD_r$ theories through the coset identity  :

\beq
\frac{SO_{k}(N)\times SO_{2}(N)}{SO_{k+2}(N)} \times\frac{SO_{1}(N)\times SO_{1}(N)}{SO_{2}(N)}=\frac{SO_{k}(N)\times SO_{1}(N)}{SO_{k+1}(N)}\times\frac{SO_{k+1}(N)\times SO_{1}(N)}{SO_{k+2}(N)}
\label{coset1}
\eeq

with 
\beq
\z{N}{p}=\frac{SO_{k}(N)\times SO_{2}(N)}{SO_{k+2}(N)} \quad p=k+N-2
\eeq

The two coset factors in the r.h.s., as well as the additional coset factor in the l.h.s. of (\ref{coset1}), correspond to $WD_{r}$ theories \cite{ref14}. For these theories the Coulomb gas representation is known. It is made of r bosonic fields, quantized with a background charge.

The equation (\ref{coset1}) can be rewritten as
\beq
\z{2r}{p}\times \WD{r}{1} = \WD{r}{k}\times \WD{r}{k+1}   \quad p=k+N-2
\label{coset2}
\eeq

This equation relates the representations of the corresponding algebras. %It could be
%reexpressed in terms of characters of representations, as is being usually done in the analyses of cosets. 
But this equation allows also to relate the conformal blocs of correlation functions. In doing so one relates the chiral (holomorphic) factors of physical operators. This later approach has been developped and analyzed in great detail in the papers \cite{ref11,ref12}, for the $SU(2)$ coset theories.

As it was said above, the chiral factor operators are related to the conformal bloc functions, not to the actual physical correlators. On the other hand, the coefficients of the operator algebra expansions are defined by the three point functions. These latter are factorizable, into holomorphic - antiholomorphic functions. So that, when the relation is established on the level of chiral factor operators, for the holomorphic three point functions, this relation could then be easily lifted to the relation for the physical correlation functions. With the relations for the chiral factor operators one should be able to define the square roots of the physical operator algebra constants.

By matching the conformal dimensions of operators on the two sides of the coset equation (\ref{coset2}) one finds the following decompositions for the operators $S$ and $A$ (here we give the results for $N \geq 8$):

\bea
S  & = & \Phi_{(111 \ldots \mid 211 \ldots)}^{(\WD{r}{k})} \times \Phi_{(211\ldots\mid 311\ldots)}^{(\WD{r}{k+1})}  \label{decompS} \\
 A & = & a\,\,\Phi_{(111\ldots\mid 111\ldots)}^{(\WD{r}{k})}\times\Phi_{(111\ldots\mid 121\ldots)}^{(\WD{r}{k+1})}
+b\,\,\Phi_{(111\ldots\mid 121\ldots)}^{(\WD{r}{k})}\times\Phi_{(121\ldots\mid 121\ldots)}^{(\WD{r}{k+1})} \label{decompA}
\eea

The Coulomb gas representation of the $\WD{r}{}$ theory allows to calculate the operator product expansion of the fields in the r.h.s. of (\ref{decompS}) and (\ref{decompA}). The coefficient $a$ and $b$ in are determined in the process, by demanding consitency between the OPE (\ref{fSS}),(\ref{fAA}),(\ref{fSA}) and the decompositions (\ref{decompS}),(\ref{decompA}) : $a=b=1/\sqrt{2}$. The following values, at the leading order in $\epsilon$, are obtained for the constants $D_1$ and $D_2$ :

\begin{eqnarray}
D_{1} & = &  2\sqrt{\frac{r}{2r-1}} \\
D_{2} & = & \frac{2(r-1)}{\sqrt{r(2r-1)}}
\end{eqnarray}

Despite the technical differences in evaluating these coefficients, in the end these results are very similar to those obtained for $N$ odd, and in both cases it boils down to : 

\begin{equation}
D_{1}=\sqrt{2}\frac{N}{\sqrt{N(N-1)}}, \quad D_{2}=\sqrt{2}\frac{N-2}{\sqrt{N(N-1)}}
\end{equation}

\begin{figure}
 \centering
 \includegraphics[scale=0.7]{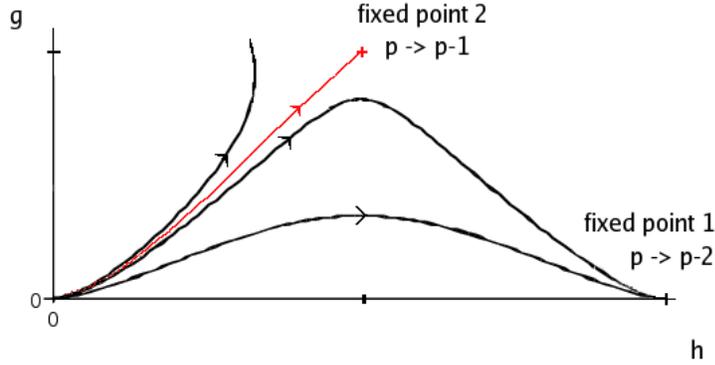}
 \caption{Renormalization group flows. Only the part $g \geq 0$ is shown, since the RG flows equations and thus the phase diagramm is symmetric under $g \rightarrow -g$.\label{phase}}
\end{figure}

At this point the analyse of the renormalization group flows becomes similar for all $N$, even or odd, and the phase diagram of constants $g$ and $h$ is essentially the same. It contains (Fig. \ref{phase}) :

\begin{itemize}
\item the trivial UV fixed point $g^{\ast}_0=h^{\ast}_0=0 $, which corresponds to the unperturbed theory $\z{N}{p}$ 

\item the standard IR fixed point on the $h$ axis: \beq (g^{\ast}_1,h^{\ast}_1) = (0,\sqrt{\frac{N(N-1)}{2}}\epsilon) \eeq described by $\z{N}{p-2}$

\item two additional IR fixed points for non-vanishing values of the two couplings, with a central charge agreeing with $\z{N}{p-1}$

\begin{eqnarray}
(g^{\ast}_2,h^{\ast}_2) & = & (\frac{1}{2}\sqrt{\frac{(N-2)(N-1)}{2}}\epsilon,\frac{1}{2}\sqrt{\frac{N(N-1)}{2}}\epsilon), \\ 
(g^{\ast}_3,h^{\ast}_3) & = & (-\frac{1}{2}\sqrt{\frac{(N-2)(N-1)}{2}}\epsilon,\frac{1}{2}\sqrt{\frac{N(N-1)}{2}}\epsilon)
\end{eqnarray}

\end{itemize}

The CFTs describing the different fixed points are identified using the $c$ function related to (\ref{potential}).

For all $N \geq 5$, the $(g,h)$ phase  diagramm is thus the same. In addition to the expected flow from $\z{N}{p}$ to $\z{N}{p-2}$, another flow to $\z{N}{p-1}$ exists. However, to realize the last flow, a fine tuning of the coupling constants $g,h$ is required. 

It is interesting to compare these results with the case $N=3$ \cite{ref17}. The perturbation of the parafermionic model $\z{3}{p}$ by two slightly relevant fields has been treated in \cite{ref11,ref12} : the RG equations admit only one IR fixed point, corresponding to the expected $\z{3}{p-4}$. No additional fixed point is present for $N=3$. Finally, the case $N=4$ is not considered here, since the corresponding second parafermionic theory can be factorized into two $N=1$ superconformal theories, for which slighlty relevant perturbations have already been studied \cite{refsuper}.

All the above analysis is valid for $p$ large, and it would be interesting to know what happens for small values of $p$.

{\bf Acknowledgements:} Very useful discussions with Vl.~S.~Dotsenko are gratefully acknowledged.

\end{document}